\input mnrass.sty
\pageoffset{-2.5pc}{0pc}

 

\Autonumber  


\pagerange{000--000}
\pubyear{1996}
\volume{000}

\begintopmatter  

\title{The effect of diffusion on the Red Giant 
luminosity function \lq{Bump}\rq }

\author{ Santi Cassisi$^{1,2}$, Scilla Degl'Innocenti$^{3,4}$ \& Maurizio Salaris$^{5,6}$}

\affiliation{$^1$Universit\'a degli studi de L'Aquila, Dipartimento di Fisica,
Via Vetoio, I-67100, L'Aquila, Italy}
\affiliation{$^2$Osservatorio Astronomico di Collurania, Via M. Maggini,
 I-64100, Teramo, Italy - E-Mail: cassisi@astrte.te.astro.it}
\affiliation{$^3$INFN, Sezione di Ferrara, Via Paradiso 12, I-44100, Ferrara, 
Italy - E-Mail: scilla@vaxfe.fe.infn.it}
\affiliation{$^4$Universit\'a degli studi di Pisa, Dipartimento di Fisica, Piazza Torricelli 2,
56100, Pisa, Italy - E-Mail: scilla@astr1pi.difi.unipi.it} 
\affiliation{$^5$Max-Planck-Institut f\"ur Astrophysik, D-85740,
Garching, Germany - E-Mail: maurizio@MPA-Garching.mpg.de} 
\affiliation{$^5$Institut d'Estudis Espacials de Catalunya,
E-08034, Barcelona, Spain}

\shortauthor{S.Cassisi, S. Degl'Innocenti \& M.Salaris}
\shorttitle{The effect of diffusion on the RGB \lq{Bump}\rq}



\abstract

\tx

This paper investigates the effect of microscopic diffusion
of helium and heavy elements on the location of the Red Giant Branch 
Luminosity Function Bump in Population II stellar models. 
To this aim updated evolutionary models taking into account 
diffusion from the Main Sequence until the Zero Age Horizontal Branch 
have been computed.  The observational luminosity difference 
between the 
RGB bump and the ZAHB ($\Delta\rm{V}^{bump}_{hb}$),
as collected for a sample of galactic globular clusters,
has been compared with the corresponding theoretical values obtained
by adopting both canonical and diffusive models. 
We find that the effect of diffusion, even if slightly improving
the agreement between observations and theory, is negligible with 
respect to the observational uncertainties. In any case the theoretical 
predictions for $\Delta\rm{V}^{bump}_{hb}$ in models with and without
diffusion appear in agreement with the observational results within the
estimated errors. Thus canonical models
 can be still safely adopted, at least until much more accurate
observational data will be available.


\maketitle  

\section{Introduction}

\tx 

From a theoretical point of a view, one of the main features of the
Red Giant Branch (RGB) of low mass stars is the existence of a region
in which the stellar luminosity, instead of monotonically increasing,
as usual, shows a significant drop before starting again to increase.
As well known (see e.g. Thomas 1967, Iben 1968, Renzini
\& Fusi Pecci 1988, Castellani, Chieffi \& Norci 1989, Bono \& Castellani 1992),
this behavior is related to the passage of the thin H burning shell
through the chemical discontinuity left by the convective envelope during the first
dredge up. Such a temporary drop in luminosity produces
a peculiar "bump" in the differential RGB luminosity function (LF),
a feature that has been  identified in the observed luminosity
functions of galactic globular clusters (GCs) ( see e.g. King et al. 1985,
Fusi Pecci et al. 1990, hereinafter FP90, Bergbush 1993, 
Sarajedini \& Norris 1994, Sarajedini \&
Forrester 1995, Brocato et al. 1996a, Brocato Castellani \& Ripepi 1995, 1996b,
Sandquist et al. 1996).

\par
For minimizing uncertainties due to the photometric data 
calibration and the cluster distance modulus, it is a common procedure,
when comparing observational RGB bump luminosities with theoretical prescriptions,
to consider the difference in the V magnitude between the RGB bump and the Zero
Age Horizontal Branch (ZAHB), i.e. the parameter $\Delta\rm{V}^{bump}_{hb}=V_{bump}-V_{hb}$.

\par
The first exhaustive comparison between theoretical and observed values 
of $\Delta\rm{V}^{bump}_{hb}$
for a large sample of GCs has been performed by FP90 by adopting
 the RGB models by Rood \& Crocker (1989) and unpublished 
HB models by Rood. Since the bump luminosity depends on the age and both the bump
and the ZAHB are affected by changes in the Helium abundance (Y)
(FP90, Cassisi \& Salaris 1997, hereinafter Paper I) FP90
explored various scenarios concerning the age and the initial Helium abundance of
galactic globular clusters.
Their conclusion was that, for constant Y, the run of the theoretical
$\Delta\rm{V}^{bump}_{hb}$ values with the metallicity was in agreement with the observations, 
but the absolute values were too bright by $\approx$0.4 mag.

\par
After the work by FP90, some suggestions have been presented in order 
to provide an
explanation of this disagreement between theoretical RGB stellar models
and observations. Some authors interpreted the discrepancy
as a proof of the intrinsic limitation of standard stellar models
and, in order to solve it, undershooting
from the formal boundary of the convective envelope (Alongi et al. 1991) has been
invoked. On the contrary, other authors (Straniero, Chieffi \& Salaris 1992, Ferraro 1992)
suggested that the discrepancy could be partially reduced simply by
computing stellar models accounting for the
observed enhancement of the $\alpha$-elements in the initial
chemical composition
of GC stars. However, for a long time this question remained unsettled.

\par
In a recent paper (Paper I) Cassisi \& Salaris investigated in detail 
this problem.
They computed theoretical models with updated equation of 
state and opacity evaluations and collected
a sample of galactic GCs with accurate photometric data
and metallicity determinations.
The dependence of the theoretical RGB bump
and ZAHB luminosity levels on different physical inputs adopted in evolutionary
computations, as for example
equation of state, opacity, mass loss, mixing length parameter, 
was analyzed.
The final result was that the agreement between observational
data and theoretical expectations was achieved within the observation 
uncertaintes, in the conservative hypothesis of coeval
GCs with the same initial Helium abundance.
Moreover it was shown that for reasonable variations of the GCs ages
and their original Helium content, the consistency with the
observations is preserved.
The discrepancy between the results in Paper I and the conclusions of FP90 
can be basically related to:
i) the use of updated stellar models ii) the adoption of new accurate spectroscopical
measurements for $[\alpha/Fe]$ and $[Fe/H]$ for GC stars iii) the accurate analysis
of recent photometric data.
The authors then concluded that {\sl there is no significant discrepancy
between observations and canonical stellar models} computed adopting updated input physics.

\par
However, during the last years, it became evident that 
the introduction of the microscopic diffusion in standard solar models
is fundamental 
to reach an agreement between the theoretical predictions of the models 
and the helioseismological observational 
results (see e.g. Christensen-Dalsgaard 1993, Guzik \& Cox 1993,  
Bahcall \& Pinsonneault 1995,
Ciacio, Degl'Innocenti \& Ricci 1997, Bahcall et al. 1996). 
Moreover, the diffusion is well known to have important consequences also
for the surface chemical composition of white dwarfs (Michaud, Fontaine \& Charland 1984).

\par
Stringfellow et al. (1983) were the first to calculate models of low mass,
metal poor stars with Helium diffusion. After this pioneering work, further 
investigations on this matter have been performed 
(Proffitt \& Vandenberg 1991, Chaboyer, Sarajedini \& Demarque 1992) with the
aim to investigate the effect of He diffusion on GCs isochrones.
More recently the effect of the diffusion of not only Helium but also of
the heavy elements on low mass, metal poor stellar models has
been taken into account by Castellani et al. (1997a). 

\par
However till now the influence of microscopic diffusion on the
$\Delta\rm{V}^{bump}_{hb}$ has been never analyzed; in this paper we 
address this topic accounting for both He and heavy elements diffusion.

\par
In Section 2 we present a brief description  
of the physical inputs and of the evolutionary 
models; the comparison with the observations and the discussion 
of the results follow
in Section 3. Finally, the conclusions are presented in Section 4.

\section{Stellar models.}
\tx
All the theoretical stellar models have
been computed by adopting the FRANEC evolutionary code (Chieffi \&
Straniero 1989) adapted
to account for the diffusion of helium and heavy elements (Ciacio et 
al. 1997). The diffusion coefficients are calculated according to
Thoul, Bahcall \& Loeb (1994). The variation of the
abundances of H, He, C, N, O and Fe due to atomic diffusion
is followed all along the structure and during all the evolutionary 
phases starting from the Zero Age Main Sequence to the RGB tip.
Following the prescription by Thoul et al. (1994),
all the other heavy elements are assumed to diffuse at the same rate as the fully
ionized iron. To account for the effect of the heavy elements
diffusion also on the opacity evaluations, we interpolated among opacity tables 
corrisponding to various metallicities (see Ciacio et al. 1997 
for more details).
\table{1}{S}{\bf Table 1. \rm Luminosities of
the ZAHB at $\log{T_{eff}}=3.85$, of the RGB bump
and the values of $\Delta\rm {V}^{bump}_{hb}$ obtained by adopting (see text) Y=0.23,
$t$=12 Gyr and using canonical stellar models.} 
{\halign{%
\rm#\hfil&\hskip10pt\hfil\rm#\hfil&\hskip10pt\hfil\rm\hfil#&\hskip10pt\hfil\rm\hfil#\cr
[M/H] & $\rm M^{zahb}_V$  & $\rm M^{bump}_V$ & $\Delta\rm {V}^{bump}_{hb}$ \cr 
\noalign{\vskip 10pt}
 -2.04 &  0.510 &  -0.198 & -0.708  \cr 
 -1.35 &  0.640 &   0.321 & -0.319  \cr
 -0.57 &  0.857 &   1.177 &  0.320  \cr}}
\par
The OPAL opacity tables (Rogers \& Iglesias
1995) for $T>10000K$ and the Alexander \& Ferguson (1994) opacities for
$T\le10000K$ 
have been used. 
The high temperature opacity tables are computed adopting the
solar heavy elements distribution by Grevesse \& Noels (1993), also
adopted in the nuclear reaction network, while the molecular opacity tables 
have been computed according to the Grevesse (1991) solar mixture. 
As for the equation of state (EOS) the OPAL EOS (Rogers,
Swenson \& Iglesias 1996) has been adopted (see also Paper I).
For this work we updated the code with respect to Paper I by
using more recent nuclear reaction 
rates for the H-burning (see Ciacio et al 1997 and Castellani et al 1997b),
the improved prescription by Haft, Raffelt \& Weiss (1994) for
the plasma neutrino energy-loss rate
(in Paper I we used the results by Munakata, Kohyama \& Itoh 1985),
and the value for the
$3\alpha$ reaction rate from Caughlan \& Fowler (1988 - in Paper I we 
used the rate from Caughlan et al 1985).

\par
As for the calibration of the superadiabatic envelope convection, we have adopted
a mixing length calibration obtained using the same procedure described in Salaris \&
Cassisi (1996). We have obtained that
mixing length ($ml$) values $ml\simeq1.6$ at Z=0.0002, $ml\simeq1.7$ 
at Z=0.001 and $ml\simeq1.8$ at Z=0.006 (coincident with the ones
obtained in Paper I for OPAL EOS models) 
have to be adopted in order to reproduce the empirical
effective temperatures of GCs RGB  as provided by Frogel et al. (1983).

\par
We evolved models of $0.70M_{\odot}$,
$0.80M_{\odot}$, $0.90M_{\odot}$ and $1.0M_{\odot}$ (fully covering 
the range of masses of stars presently evolving along the RGB of GCs)
with Z=0.0002, Z=0.001 and Z=0.006 and an initial
helium abundance equal to Y=0.23 from the Zero Age Main
Sequence to the RGB tip.
For each fixed metallicity, we interpolated among these models to obtain 
the value of $V_{bump}$ (as defined in Paper I) for an age of t=12Gyr.

\par
This age (also adopted in Paper I) has been chosen 
following the recent results of
different groups about the age of galactic GCs
(Chaboyer \&
Kim 1995, Mazzitelli, D'Antona \& Caloi 1995, Salaris, Degl'Innocenti
\& Weiss 1997).
As discussed by Castellani et al. (1997a), the combined effect of He and heavy
elements diffusion reduces the estimated ages of GCs by only less than 1 Gyr.

\par
Following the same procedure adopted in Paper I, for each chemical
composition we computed also ZAHB models. 
For the transformation from the theoretical to the observational plane we 
used, as in Paper I, the
Kurucz (1992) transformations, adopting $M_{Bol,\odot}=4.75$ mag.
Then the theoretical value of $\Delta\rm{V}^{bump}_{hb}$ has been obtained by considering
the difference between the V magnitudes of the RGB bump and of the ZAHB taken at
$\log{T_{eff}}=3.85$, which can be safely adopted as the average temperature of the
RR Lyrae instability strip.
This procedure has been followed for both canonical stellar models and
for the models which take into account microscopic diffusion.
The bump and the ZAHB V magnitudes and the value of 
$\Delta\rm{V}^{bump}_{hb}$ which refer to the canonical stellar models (computed with
the physical inputs discussed above), are shown in Table 1, while the same quantities 
for the models with He and heavy elements diffusion are
reported in Table 2. As in Paper I, we define
$\rm [M/H]=\log(M/H)_{star}-\log(M/H)_{\odot}\approx\log(Z)-1.65$.

\table{2}{S}{\bf Table 2. \rm As in Table 1, but in this case 
for the stellar models
which include microscopic diffusion of helium and heavy elements.} 
{\halign{%
\rm#\hfil&\hskip10pt\hfil\rm#\hfil&\hskip10pt\hfil\rm\hfil#&\hskip10pt\hfil\rm\hfil#\cr
[M/H] & $\rm M^{zahb}_V$  & $\rm M^{bump}_V$ & $\Delta\rm {V}^{bump}_{hb}$ \cr 
\noalign{\vskip 10pt}
 -2.04 &  0.548 &  -0.129 & -0.677  \cr 
 -1.35 &  0.661 &   0.392 & -0.269  \cr
 -0.57 &  0.872 &   1.257 &  0.385  \cr}}

\par
Figure 1 shows the values of 
$\Delta\rm{V}^{bump}_{hb}$ as a function of the global amount of heavy elements when
the microscopic diffusion is properly included in stellar computations
(the lines have been obtained by computing the parabolic
function which crosses all points). 
In order to evaluate easily the effect of the diffusion,
we have also plotted the theoretical relation obtained adopting 
the same physical inputs
but canonical stellar models (hereinafter we refer to it as {\sl new canonical scenario}).
For the sake of comparison, the theoretical relation corresponding to the standard stellar
models computed adopting the OPAL EOS, as presented in
Paper I (corresponding to the same average age t=12 Gyr),
is also displayed ({\sl old canonical scenario}). 
\figure{1}{S}{80mm}{\bf Figure 1. \rm 
Theoretical values of $\Delta\rm{V}^{bump}_{hb}$ $versus$ the global
metallicity as obtained using stellar models computed accounting for 
microscopic diffusion. The theoretical prescriptions obtained by using 
standard stellar models, and the canonical relation presented in Paper I
are also displayed.}

\par
Before discussing the net effect on $\Delta\rm{V}^{bump}_{hb}$ due to the
inclusion of microscopic diffusion in stellar computations, it is important
to notice the difference, in particular at low metallicities,
between the {\sl \lq{old}\rq} and {\sl \lq{new}\rq\ canonical scenario}. 
Let us remind that the only differences between the two sets of canonical stellar models
adopted for computing these theoretical relations are related to the nuclear reaction rates
for the H-burning, the $3\alpha$ reaction rate and the plasma neutrino energy-loss rate.
The main contribution to the changes of $\Delta\rm{V}^{bump}_{hb}$ between the {\sl old} and
{\sl new} canonical scenario comes from the different luminosity of the ZAHB, since the different 
core mass at the RGB tip due to the new plasma neutrino energy-loss rate
and the $3\alpha$ reaction rate.
This difference between the two standard scenarios can be assumed as
a crude estimate of the uncertainty of the standard
theoretical $\Delta\rm{V}^{bump}_{hb}$ values
related to uncertainties in 
neutrino energy losses and energy generation rates.

\par
A more detailed discussion about the stellar models presented 
in this paper and the differences between the {\sl old} and {\sl new}
canonical scenario, a comparison with previous computations 
and an analysis of the effects of the inclusion of He and heavy elements
diffusion on both 
isochrones and He burning models can be found in Cassisi et al. (1997).

\par
To investigate the effect of the diffusion,
we have to compare the non-standard evolutionary scenario with the canonical
one obtained using the same physical inputs ({\sl new canonical scenario}).
The result is that the proper inclusion of both He and heavy elements
diffusion produces slightly larger $\Delta\rm{V}^{bump}_{hb}$ value
with respect to the new canonical scenario (see Table 2);
the difference is on average by 0.05 mag and at maximum 
only by $\approx$0.065 mag (at $[M/H]$=-0.57).
This result is in agreement with the early 
suggestion provided at the end of Paper I
and it can be understood when taking into account the following evidences:
\medskip
\noindent
i) for each fixed stellar mass and chemical composition, when diffusion is 
taken into account, the RGB bump is located
at lower luminosity ($\Delta\rm V_{bump}\approx0.07$ mag, the exact value
depending on the chemical composition). This effect is related to
the higher opacity at the base of the 
convective envelope, owing to the decrease of the He
abundance (and the corresponding increase of the hydrogen abundance) in the envelope;
\smallskip
\noindent
ii) the luminosity level of the ZAHB is also decreased 
($\Delta\rm V_{ZAHB}\approx0.02$ mag on average,
the exact value depending again on the metallicity),
due to the lower amount of Helium in the
envelope when the stars settles on the Horizontal Branch, that causes 
a lower efficiency of the H burning shell;
\smallskip
\noindent
iii) the second effect partially compensates the first one, so that 
the inclusion of microscopic diffusion in stellar computations does not modify significantly
the theoretical scenario for the relation between $\Delta\rm{V}^{bump}_{hb}$ and the 
metallicity.

\section{Comparison with the observations.}

\tx

As discussed in Paper I, for performing a meaningful comparison between theoretical and 
observed $\Delta\rm{V}^{bump}_{hb}$ values one needs to select a sample
of GCs for which high quality photometric data and accurate spectroscopical
measurements of $[Fe/H]$ and $[\alpha/Fe]$ are available.
Following this prescription we selected
a sample of seven GCs with a quite large range of heavy elements 
abundances: from a
metal poor cluster as NGC6397 to the more metal rich ones as
47Tuc. We refer to Paper I for the description of the procedure adopted
to obtain the ZAHB and RGB bump luminosity levels and for the sources of the
photometric data and spectroscopical metallicities. 
We report in Table 3 only the data needed in order to compare models 
and observations.
As for the relation between the global amount of heavy elements and the
values of $[Fe/H]$ and $[\alpha/Fe]$ we adopt, as in Paper I,
the relation (see Salaris et al. 1993 for more details):
$$[M/H]\cong[Fe/H]+\log(0.638\cdot{f}+0.362) $$
where $\log(f)=[\alpha/Fe]$ is the enhancement factor of the $\alpha$ elements.
This relation has been derived by using, as the reference solar heavy element
distribution, the Ross \& Aller (1976) mixture. However, we have verified
that an analogous relation derived (following the prescriptions by
Salaris et al.\ 1993) by using the Grevesse \& Noel (1993) solar metal
distribution provides $[M/H]$ values different by not more than
0.01 dex in the range of $[\alpha/Fe]$ we are dealing with.
The global metallicities adopted in Paper I 
(from Salaris \& Cassisi 1996) are reported in column 5
of Table 3.

\par
It is also worth noting that recently Carretta \& Gratton (1997) have provided a new
compilation of $[Fe/H]$ values for a large sample of galactic GCs.
For the clusters 47Tuc, NGC6752 and NGC6397, Carretta \& Gratton have
used new high resolution spectra, while for the other clusters in their sample, they have
adopted high quality literature data (namely, equivalent widths from high dispersion spectra).
All these data have been analized with an homogeneous and self-consistent procedure by adopting
the model atmospheres by Kurucz (1992).

\par
Since the $[Fe/H]$ values
provided by Carretta \& Gratton are systematically larger than the values
given by Gratton and coworkers in previous works and largely adopted by Salaris \& 
Cassisi (1996) in their compilation of galactic GC  metallicities, it is
interesting to test what is the effect of this new metallicity scale on the present investigation.
For this reason, we report also in Table 3 (column 6), the value of $[M/H]$ obtained by adopting
for each cluster the value of $[Fe/H]$ provided by Carretta \& Gratton, but still relying
on the $[\alpha/Fe]$ values reported in Salaris \& Cassisi (1996).

\par
The [M/H] values from Paper I are between 0.05 and 0.16 dex lower
than the ones derived using the Carretta \& Gratton  [Fe/H] scale, with the
exception of M5; in this case the Carretta \& Gratton metallicity is larger by 0.29 dex
 with respect to the values from Paper I.

\table{3}{S}{\bf Table 3. \rm Visual magnitude of the bump and
of the ZAHB at the RR Lyrae instability strip, and $\Delta\rm {V}^{bump}_{hb}$
for a sample of galactic GCs (see Paper I). The last two columns list the global amount of
heavy elements as obtained by adopting the $[Fe/H]$ values given in Salaris \& Cassisi (1996)
and in Carretta \& Gratton (1997), respectively.} 
{\halign{%
\rm#\hfil&\hskip3pt\hfil\rm#\hfil&\hskip4pt\hfil\rm#\hfil&\hskip4pt\hfil\rm\hfil#\hfil&\hskip4pt\rm\hfil#\hfil&\hskip5pt\rm\hfil#\hfil\cr
NGC & $\rm V_{zahb}$  & $\rm V_{bump}$ & $\Delta\rm {V}^{bump}_{hb}$ & $\rm [M/H]_{SC}$ & $\rm [M/H]_{CG}$ \cr 
\noalign{\vskip 10pt}
   104 (47Tuc) & 14.20 &  14.55 &  $\,\,\,\,0.35\pm0.18$ & -0.70 & -0.60 \cr
  1904 (M79) & 16.36 &  16.00 &  $-0.36\pm0.12$ & -1.27 & -1.22 \cr
  5272 (M3)  & 15.76 &  15.40 &  $-0.36\pm0.07$ & -1.31 & -1.16 \cr
  5904 (M5)  & 15.15 &  14.95 &  $-0.20\pm0.07$ & -1.19 & -0.90 \cr
  6352        & 15.50 &  15.86 &  $\,\,\,\,0.36\pm0.12$ & -0.70 & -0.54  \cr
  6397        & 13.02 &  12.60 &  $-0.42\pm0.14$ & -1.70 & -1.64 \cr
  6752        & 13.86 &  13.65 &  $-0.21\pm0.12$ & -1.28 & -1.20 \cr}}

Figure 2a shows the comparison between the observational values of  
$\Delta\rm {V}^{bump}_{hb}$ for the selected clusters and the theoretical
prescriptions for standard stellar models ({\sl new canonical
scenario}) and for models evolved accounting for He and heavy elements diffusion. 
The theoretical relations obtained adopting stellar models with
microscopic diffusion for two different cluster
ages are also plotted. The clusters metallicities are the ones used
in Paper I (here, as in Figure 2b, an error bar on the $[M/H]$ values
by $\pm$0.15 dex is assumed).
\figure{2}{S}{120mm}{\bf Figure 2. \rm {\it a})
The values of $\Delta\rm{V}^{bump}_{hb}$ $versus$ the global
metallicity for all clusters selected in Paper I.
Our theoretical relation obtained using stellar models computed accounting for 
microscopic diffusion (and three different assumptions concerning the cluster age) is
displayed, together with the relation obtained by using standard stellar models;
{\it b}) as in Panel {\it a}) but adopting the Carretta \& Gratton (1997) metallicity
scale (see text for more details).}
One notices that the agreement between the theoretical and
observational values of $\Delta\rm{V}^{bump}_{hb}$, obtained by
adopting the {\sl new canonical
scenario}, is only slightly improved by the introduction of the
microscopic diffusion in the stellar models. As a matter of fact
the effect of diffusion is not relevant with 
respect to the observational uncertaintes. 
Moreover, as already found in Paper I, the present result is not
significantly modified by relaxing the hypothesis of globular clusters coeval
and 12 Gyr old.  This is shown, in the same Figure 2a, by the location of the two
theoretical lines corresponding to  diffusive models for
two different assumptions about the age of the stellar systems.

\par 
In Figure 2b the same observational data and theoretical results are
plotted, but this time the clusters $[M/H]$ are derived by
adopting the Carretta \& Gratton  $[Fe/H]$ scale.
The agreement between theory and observations is slightly worst, also
if again theory and observations agree within the observational uncertainties and 
the indetermination on the GCs ages. 
There is only the observational point corresponding to M5 that is clearly not
reproduced by the theoretical values and, it is important to note that M5
is also the cluster for which the Carretta \& Gratton scale 
shows the largest difference with respect to previous $[Fe/H]$ estimates.

\par
Since, as clearly stated by Carretta \& Gratton (1997), 
the reason for the difference between their
$[Fe/H]$ scale and previous determinations is mainly due
to the use of the Kurucz (1992) model atmospheres, whose reliability
should be tested by independent computations,
we tend to consider the difference between the values of $[M/H]$ reported 
in columns 5 and 6 of Table 3 as an estimate of the uncertainty on the 
spectroscopic $[M/H]$ determinations presently available.

\par
In conclusion, more numerous and much more precise observational data 
are needed before obtaining a clear evidence about the
greater reability of a theoretical scenario in comparison with the other 
ones.

\section{Final remarks.}

\tx
To investigate the effect of microscopic diffusion
on the RGB Luminosity Function Bump we computed updated stellar models of low mass, 
metal poor stars, in the
canonical scenario and by including helium and heavy elements diffusion,
from the Zero Age Main Sequence to the ZAHB phase.

\par
As a result, we found that the values of $\Delta\rm{V}^{bump}_{hb}$ obtained 
for stellar models with diffusion are only slightly
larger than the canonical ones (the maximum difference is $\approx$0.06 
mag), both being in general good agreement with 
the corresponding observational values.

\par
Due to this very small influence of the diffusion on the $\Delta\rm{V}^{bump}_{hb}$ values,
especially if compared with the typical uncertainties related to the  observational
determinations of [M/H] and $\Delta\rm {V}^{bump}_{hb}$, 
{\sl stellar standard models
can be still safely adopted in analyzing this important diagnostic of the
inner chemical stratification in
low mass stars}, at least until much more accurate
data for a larger sample of GCs will be available. 

\par
However, we point out 
that at present time
the introduction of the Helium and heavy elements
diffusion in stellar computations seems to be
a "forced stage" in the analysis of
helioseismological data  or of the 
surface chemical abundances of white dwarfs and
extremely hot horizontal branch stars 
(Moehler et al. 1995 and reference therein).

\section*{Acknowledgments}

\tx 

We warmly thank F. Ciacio for providing us with the subroutine for the
calculation of microscopic diffusion of helium and heavy elements.
We are very grateful to V. Castellani for his continuous encouragement
and for useful discussions as well as for reading a preliminary
draft of the paper. We thank also the referee for her/his
useful suggestions.

\section*{References}

\bibitem Alexander D.R. \& Ferguson J.W. 1994, ApJ 437, 879
\bibitem Alongi M., Bertelli G., Bressan A. \& Chiosi C. 1991, A\&A 244, 95
\bibitem Bahcall J.N. \& Pinsonneault M.H. 1995, Rev.Mod.Phys. 76, 781
\bibitem Bahcall J.N., Pinsonneault M.H., Basu S. \& Christensen-Dalsgaard J. 1996,
preprint IASSNS-AST 96/54
\bibitem Bergbush P.A. 1993, AJ 106, 1024
\bibitem Bono G. \& Castellani V. 1992, A\&A 258, 385
\bibitem Brocato E., Buonanno R., Malakhova Y. \& Piersimoni A.M. 1996a, A\&A 311, 778
\bibitem Brocato E., Castellani V. \& Ripepi V. 1995, AJ 109, 1670
\bibitem Brocato E., Castellani V. \& Ripepi V. 1996b, AJ 111, 809
\bibitem Carretta E. \& Gratton R.G. 1997, A\&AS 121, 95
\bibitem Cassisi S., Castellani V., Degl'Innocenti S. \& Weiss
A. 1997, A\&A submitted
\bibitem Cassisi S. \& Salaris M. 1997, MNRAS, 285, 593
\bibitem Castellani V., Chieffi A. \& Norci L. 1989, A\&A 216, 62
\bibitem Castellani, V., Ciacio F., Degl'Innocenti S. \& Fiorentini
G. 1997a, A\&A in press
\bibitem Castellani, V., Degl'Innocenti, S., Fiorentini, G., Lissia, M. 
\& Ricci, B. 1997b, preprint INFNFE-10-96, to appear on Phys. Rep.
\bibitem Caughlan, G.R. \& Fowler, W.A. 1988, Atom.Data Nucl. Data Tables 40, 283
\bibitem Caughlan, G.R., Fowler, W.A., Harris, M.J. \& Zimmermann, B.A. 
1985, Atomic Data \& Nuclear Data Tables, 32, 197
\bibitem Chaboyer B. \& Kim Y.-C. 1995, ApJ 454, 767
\bibitem Chaboyer D., Sarajedini A. \& Demarque P. 1992, ApJ 394, 515
\bibitem Chieffi A. \& Straniero O. 1989, ApJS 71, 47
\bibitem Ciacio F., Degl'Innocenti S. \& Ricci B. 1997, A\&A in press
\bibitem Christensen-Dalsgaard J., Proffit C.R. and Thompson M.J. 1993, 
ApJ 403, L75
\bibitem Guzik J.A. \& Cox A.N. 1993, ApJ 411, 394
\bibitem Ferraro F.R. 1992, MemSAIt 63, 491
\bibitem Frogel J.A., Persson S.E. \& Cohen J.G. 1983, ApJS 53, 713
\bibitem Fusi Pecci F., Ferraro F.R., Crocker D.A., Rood R.T. \& 
Buonanno R. 1990,A\&A 238,95
\bibitem Grevesse N. \& Noels A. 1993, in "Origin and Evolution of the elements",
eds. Prantzos N., Vangioni-Flam E., Casse M. (Cambridge Univ. Press, Cambridge), P. 15
\bibitem Haft, M., Raffelt, G. \& Weiss, A. 1994, ApJ, 425, 222
\bibitem Iben I. Jr 1968, Nature 220, 143
\bibitem Itoh, N.,Adachi, T., Nakagawa, M., Kohyama, Y. \& Munakata, H. 1989,
ApJ, 339, 354; erratum 360, 741 (1990)
\bibitem King C.R., Da Costa G.S. \& Demarque P. 1985, ApJ 299, 674
\bibitem Kurucz R.L. 1992, in Barbuy B., Renzini A. (eds.), IAU
Symp. n. 149, ``The Stellar Populations of Galaxies'', Kluwer,
Dordrecht, p. 225
\bibitem Mazzitelli I., D'Antona F. \& Caloi V. 1995, A\&A 302, 382
\bibitem Michaud G., Fontaine G. \& Charland Y. 1984, ApJ, 280, 787 
\bibitem Moehler S., Heber U. \& De Boer K.S. 1995, A\&A 294, 65
\bibitem Munakata, H., Kohyama, Y. \& Itoh, N. 1985, ApJ 296, 197
\bibitem Proffitt C.R. \& Vandenberg D.A. 1991, ApJS 77, 473
\bibitem Renzini A. \& Fusi Pecci F. 1988, ARA\&A 26, 199
\bibitem Rogers F.J. \& Iglesias C.A. 1995,
in Adelman S.J., Wiesse W.L. (eds.), "Astrophysical application of powerful
new databases", ASP Conference Series, vol. 78, p.31
\bibitem Rogers F.J., Swenson F.J. \& Iglesias C.A. 1996, ApJ 456, 902
\bibitem Rood R.T. \& Crocker D.A. 1989, in Schmidt E.G. (ed.), IAU
Coll. 111, ``The use of Pulsating stars in Fundamental Problems of 
Astronomy'', Cambridge University Press, p. 103
\bibitem Ross J.E. \& Aller L.H. 1976, Science 1991, 1223
\bibitem Salaris M. \& Cassisi S. 1996, A\&A 305, 858
\bibitem Salaris M., Degl'Innocenti S. \& Weiss A. 1997, ApJ in press
\bibitem Sandquist E.L., Bolte M., Stetson P.B. \& Hesser J.E. 1996, ApJ 470, 910
\bibitem Sarajedini A. \& Norris J.E. 1994, ApJS 93, 161
\bibitem Sarajedini A. \& Forrester W.L. 1995, AJ 109, 1112
\bibitem Straniero O., Chieffi A. \& Salaris M. 1992, MemSAIt 63, 315
\bibitem Stringfellow G.S., Bodenheimer P., Noerdlinger P.D. \& Arigo R.J. 1983
ApJ 264, 228
\bibitem Thomas H.-C. 1967, Z.Ap. 67, 420
\bibitem Thoul A.A., Bahcall J.N. \& Loeb A. 1994, ApJ 421, 828

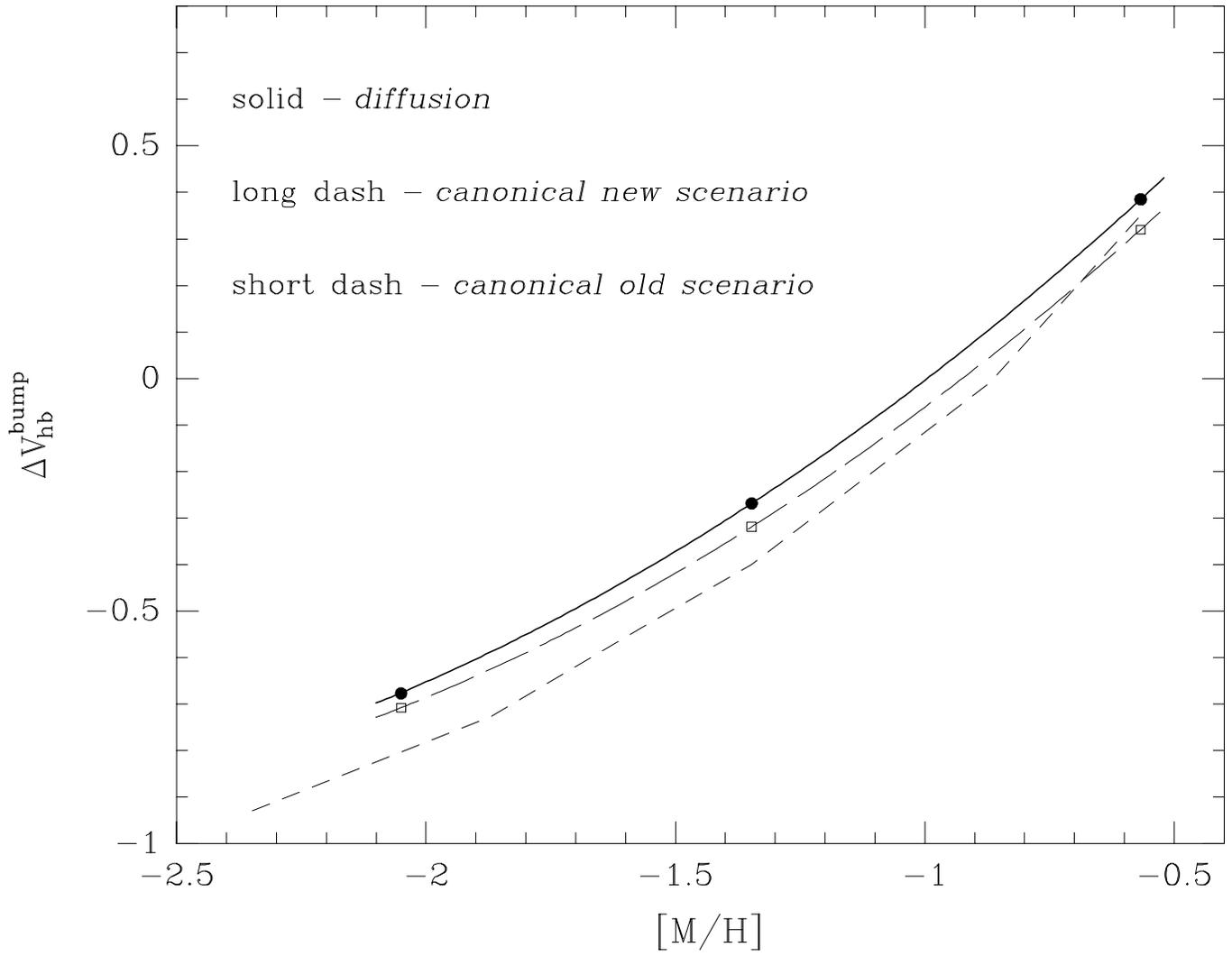
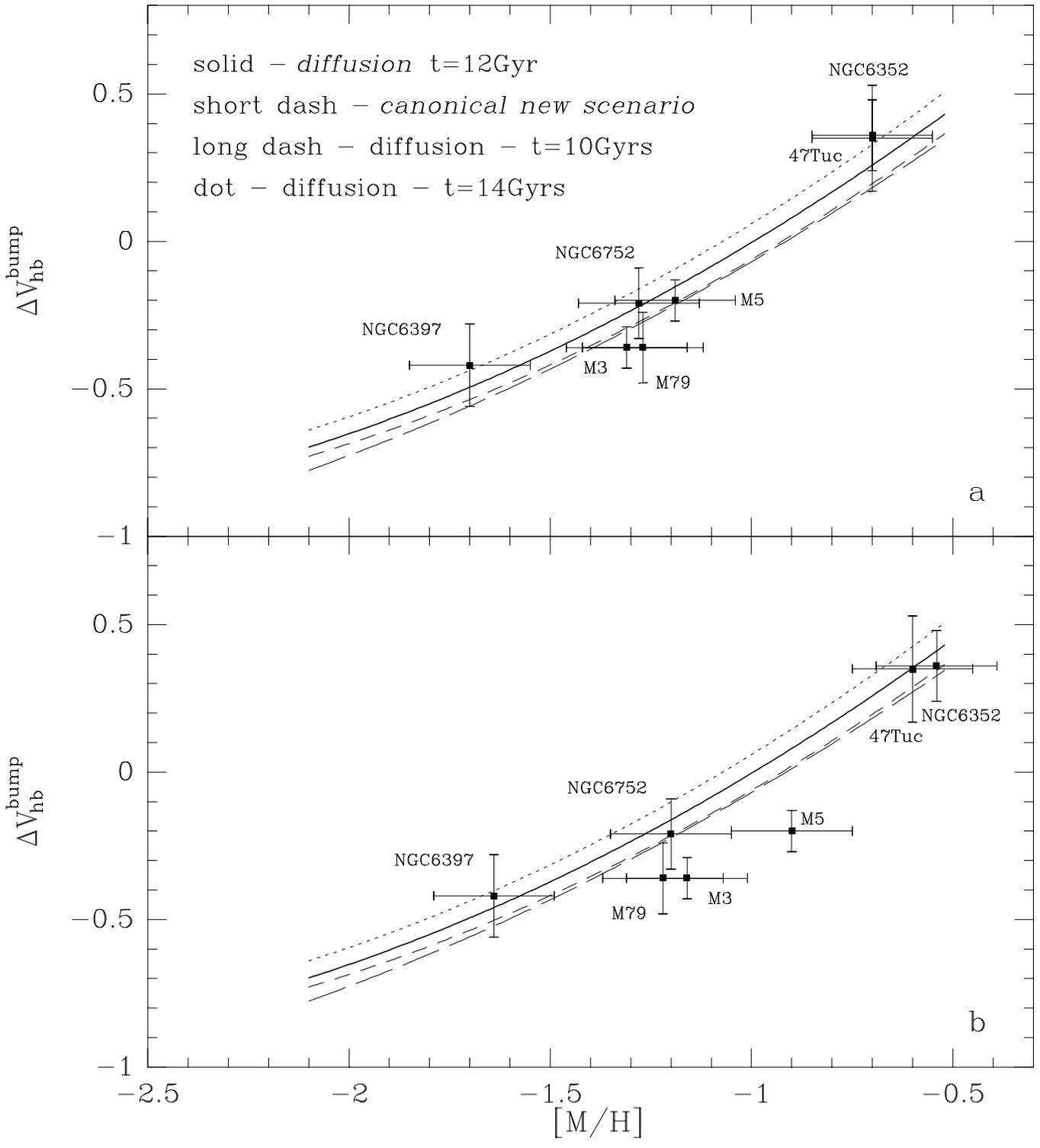
\bye